\documentclass[preprint,authoryear,14pt,graphics]{aastex}
\usepackage{graphics}
\usepackage{amssymb}
\begin{document}
%% Title, authors and addresses
%% use the tnoteref command within \title for footnotes;
%% use the tnotetext command for the associated footnote;
%% use the fnref command within \author or \address for footnotes;
%% use the fntext command for the associated footnote;
%% use the corref command within \author for corresponding author footnotes;
%% use the cortext command for the associated footnote;
%% use the ead command for the email address,
%% and the form \ead[url] for the home page:
%%
%% \title{Title\tnoteref{label1}}
%% \tnotetext[label1]{}
%% \author{Name\corref{cor1}\fnref{label2}}
%% \ead{email address}
%% \ead[url]{home page}
%% \fntext[label2]{}
%% \cortext[cor1]{}
%% \address{Address\fnref{label3}}
%% \fntext[label3]{}

\title{White Paper for Blazar Observations with a GEMS-like X-ray Polarimetry Mission}
\author{
Henric Krawczynski\altaffilmark{1},
Lorella Angelini\altaffilmark{2},
Matthew Baring\altaffilmark{3},
Wayne Baumgartner\altaffilmark{2},
Kevin Black\altaffilmark{2},
Jessie Dotson\altaffilmark{4},
Pranab Ghosh\altaffilmark{5},
Alice Harding\altaffilmark{2},
Joanne Hill\altaffilmark{2},
Keith Jahoda\altaffilmark{2},
Phillip Kaaret\altaffilmark{6},
Tim Kallman\altaffilmark{2},
Julian Krolik\altaffilmark{7},
Dong Lai\altaffilmark{8},
Craig Markwardt\altaffilmark{2},
Herman Marshall\altaffilmark{9},
Jeffrey Martoff\altaffilmark{10},
Robin Morris\altaffilmark{4},
Robert Petre\altaffilmark{2},
Juri Poutanen\altaffilmark{11},
Stephen Reynolds\altaffilmark{12},
Jeffrey Scargle\altaffilmark{4},
Jeremy Schnittman\altaffilmark{2},
Peter Selemitsos\altaffilmark{2},
Yang Soong\altaffilmark{2},
Tod Strohmayer\altaffilmark{2},
Jean Swank\altaffilmark{2},
and Toru Tamagawa\altaffilmark{14}}

\altaffiltext{1}{Department of Physics, Washington University in St. Louis}
\altaffiltext{2}{NASA/GSFC}
\altaffiltext{3}{Department of  Physics and Astronomy, Rice University}
\altaffiltext{4}{NASA/ARC}
\altaffiltext{5}{Department of Astronomy and Astrophysics, TIFR, Mumbai, India}
\altaffiltext{6}{Department of Physics and Astronomy, University of  Iowa}
\altaffiltext{7}{Department of Physics and Astronomy, Johns Hopkins University}
\altaffiltext{8}{Department of Astronomy, Cornell University}
\altaffiltext{9}{Center for Space Research, MIT}
\altaffiltext{10}{Department of Physics, Temple U.}
\altaffiltext{11}{U. Oulu, Finland}
\altaffiltext{12}{Department of Physics and Astronomy, North Carolina State University}
\altaffiltext{14}{Riken University, Japan}

\begin{abstract}
In this document, we describe the scientific potential of blazar observations 
with a X-ray polarimetry mission like GEMS (Gravity and Extreme Magnetism SMEX). 
We describe five blazar science investigations that such a mission would enable:
(i) the structure and the role of magnetic fields in AGN jets,
(ii) analysis of the polarization of the synchrotron X-ray emission from AGN jets,
(iii) discrimination between synchrotron self-Compton and external Compton
models for blazars with inverse Compton emission in the X-ray band,
(iv) a precision study of the polarization properties of the X-ray 
emission from Cen-A,
(v) tests of Lorentz Invariance based on X-ray polarimetric 
observations of blazars.
We conclude with a discussion of a straw man observation program and recommended 
accompanying multiwavelength observations.
\end{abstract}
% \linenumbers
%% main text
\section{Introduction}
\label{intro}
Blazars are Active Galactic Nuclei (AGN) with jets (highly relativistic outflows) 
that are aligned with the line of sight to within a few degrees. 
Owing to the effect of relativistic beaming, blazars are among the 
brightest extragalactic X-ray sources and allow us to study their 
non-thermal jet emission with excellent signal-to-noise ratios \citep{Kraw:13}. 

X-ray polarimetry is still an emerging field. 
NASA has launched so far only one satellite-borne instrument specifically designed for X-ray polarimetry, 
the Graphite Crystal X-ray polarimeter experiment on the {\it OSO-8} mission.
The polarimeter detected polarized X-ray emission from the Crab Nebula  \citep{Weis:78} but
was not sensitive enough to detect polarization from extragalactic X-ray sources. 
An X-ray polarimetry mission like GEMS (Gravity and Extreme Magnetism SMEX) can access 
the polarization of the 2-10 keV emission of a 1 mCrab source down to polarization degrees of 
2.8\% MDP (99\% confidence level Minimum Detectable Polarization for a 10$^6$~s observation) \citep{Blac:10}. 
GEMS was proposed in response to the NASA SMEX announcement of opportunity in December 2008, was selected for 
phase A development in 2009 and selected for phase B in 2010.
Unfortunately, the mission was not confirmed in 2012 owing to
the risk of a cost overrun. 

A mission like GEMS has the potential to make a number of smoking-gun 
observations concerning the inner workings of AGN jets: 
\begin{enumerate}
\item How is the magnetic field in AGN jets structured, and which role does it play 
in launching the jet, accelerating it, and confining it? 
\item Is the X-ray emission from high-energy peaked BL Lac objects  
indeed synchrotron emission? If yes, is the radio, IR, or optical radiation 
from the AGN core emitted co-spatially with the X-ray emission?   
\item Is the X-ray emission from Flat Spectrum Radio 
Quasars (FSRQs) and low and intermediate energy peaked BL Lac objects indeed 
inverse Compton emission, and what is the dominant target photon population 
for the inverse Compton processes? 
\item How are the properties of the sub-pc scale jet (from which the
core X-ray emission comes) related to the properties of the kpc-scale jets?
\item Is Lorentz Invariance a low-energy phenomenon?
\end{enumerate}
The objective of this white paper is to describe these science drivers of a GEMS-like 
X-ray polarimetry mission that operates in the 2-10 keV energy range 
and to identify measurements that have a high likelihood to make 
breakthrough discoveries. Furthermore, the document describes the 
observation strategy and list preparations required to maximize 
the science return. We describe specific science investigations 
in Sect.\ \ref{gun}. For each investigation, the science drivers 
are explained, a list of the best targets is provided, and 
the required accompanying multiwavelength coverage is discussed.   
We describe a straw man observation program in Sect.\ \ref{strategy}, and comment
on the termination of the GEMS mission in Sect.~\ref{gems}.

One distinguishes between blazars according to the frequency at which the Spectral 
Energy Distribution (SED) of the low-energy (synchrotron) component peaks. 
In the following we use the abbreviations LSP, ISP and HSP to denote low synchrotron 
peaked ($\nu_{\rm peak}^{\rm S}<10^{14}$~Hz), 
intermediate synchrotron peaked ($10^{14}$~Hz$<\nu_{\rm peak}^{\rm S}<10^{15}$~Hz), and
high synchrotron peaked ($\nu_{\rm peak}^{\rm S}>10^{15}$~Hz) blazars.
In the earlier literature, the reader will often find the abbreviations LBL, IBL, and HBL.
These names denote the BL Lac subclasses of LSP, ISP, and HSP blazars, respectively.  
\section{Blazar Science Investigations}
\label{gun}
\subsection{The Structure and Role of Magnetic Fields in AGN Jets}
\label{structure}
\citet{Mars:08} and \citet{Abdo:10} reported tentative 
evidence for optical polarization swings that were associated with $\gamma$-ray flares.
The authors interpret the observations in the framework of a model in which AGN jets
are accelerated and confined by helical magnetic fields. A jet segment moves through a
stationary shock, and particle acceleration at this shock gives rise to the observed 
flaring episodes. As the helical magnetic field of the jet moves 
through the standing shock the polarization of the observed emission 
exhibits a swing.

A mission like GEMS has the potential to reveal similar swings in the X-ray regime. 
The electrons responsible for the X-ray synchrotron emission cool faster than the electrons
responsible for the optical synchrotron emission. The X-ray emission regions are probably 
much smaller and thus more uniform than the optical emission regions.
X-ray polarization observations may thus deliver cleaner evidence for the helical
magnetic field structure at the base of AGN jets than optical polarization observations. 
The GEMS energy band is well suited to detect the synchrotron emission of HSPs.
Only very few HSPs exhibit readily detectable synchrotron emission at even higher energies. 
The X-ray observations might detect higher degrees of polarization and/or more frequent 
polarization swings. If the X-ray polarization indeed shows frequent polarization swings, 
the observations could prove the helical magnetic field structure at the base of AGN jets, 
and could thus make a key contribution to our understanding of the formation, acceleration, 
and collimation of jets.

\begin{figure}[tb]
\begin{center}
\includegraphics[width=4.5in]{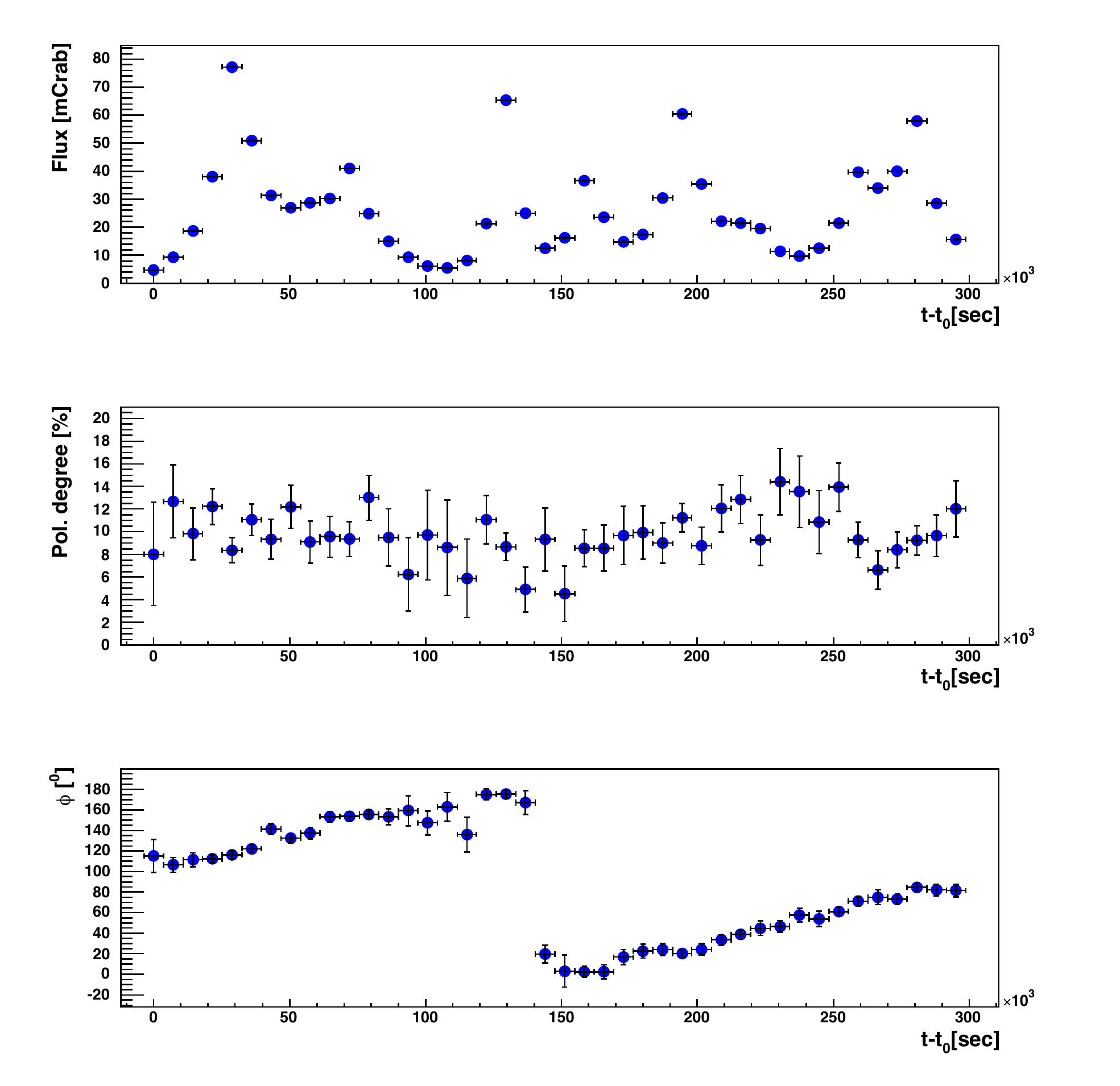}
\caption{\label{m421} Simulation of a 300 ksec GEMS observation of a Mrk 421 flaring epoch.
The simulations assume flux levels between 0 and 80 mCrab (upper panel), and a constant
polarization degree of 10\% (center panel). The polarization exhibits a swing with a rate of
$d\phi/dt=$ 50$^{\circ}$.}
\end{center}
\end{figure}
Figure \ref{m421} shows the results of a simulated 300 ksec (80 hrs) GEMS 
observations of the HSP Mrk 421. For over a decade the source has been one of 
the X-ray brightest HSPs in the sky. It is a prime target for GEMS
as it exhibits high polarization degrees of about $\sim$15\% in the optical \citep{Sill:93}. 
The data in Fig.\ \ref{m421} are shown for 2 hrs bins, assuming that the source is visible 
for 80\% of the time. The calculations assume that the polarization degree of the 
synchrotron X-ray emission is similar to the polarization degree of the optical emission.

In the case of the source BL Lac, \citet{Mars:08} measured a swing of the 
optical polarization with a rate $d\phi/dt=$ 50$^{\circ}$ per day. 
Our simulation assumed that the X-ray polarization of Mrk 421 behaves in a similar way
as the optical polarization of BL Lac as both sources exhibit similar optical
polarization degrees \citep{Sill:93,Tost:98}. Furthermore, the galactic bulge black 
hole mass estimates are similar for the two sources: 
$\approx 2\times$ 10$^8$ M$_{\rm sol}$ \citep{Bart:03,Mars:08}.
We thus expect a similar $d\phi/dt$ for the two sources as the swing rate 
should scale with the mass of the black hole.

Good targets are X-ray bright and show a high degree of optical polarization.
Four X-ray bright objects from the compilation of HSPs and ISPs of 
\citet{Cost:02} are given in Table 1 plus 1ES 1218+304. The table also lists 
the polarization degree measured in the optical band. 
Note that blazars exhibit flares with a ``red noise'' power spectrum. 
This means that the flux levels change not only on short time scales of a few minutes but also on time scales of weeks, months and years. 
Ideally, the observation program should be adapted to up-to-date information about 
flaring sources. The best strategy will combine archival information (as the one from Table 1) with
source activity indicators from ground based and space borne satellites.

\begin{table}[t]
\begin{center}
\begin{tabular}{|p{2.4cm}|p{1.2cm}|p{1.2cm}|p{2.4cm}|p{2.4cm}|}
\hline
Object			& Class & $z$   & $F(2-10\,\rm keV)$ & opt.\ pol $\left[ \% \right]$  \\ \hline \hline
Mrk 421			& HSP   & 0.031 & 48    & 0-13 (1,2)		 \\ 
Mrk 501			& HSP   & 0.034 & 24    & 2-4 (1)		  \\
1ES 1959+650	& HSP	& 0.047 & $<$12 & nd			  \\
PKS 2155-314	& HSP   & 0.116 & $<$12 & 2-10 (2,3,4)	  \\
1ES 1218+304    & HSP	& 0.184 & 24    & 5  (3)		  \\ \hline
\end{tabular}
\end{center}
\caption{\label{tflux}  \small Target list for the science investigations  
``The Structure and Role of Magnetic Fields in AGN Jets''
and the science investigation ``Study of the Synchrotron X-ray Emission from AGN Jets''.
Column 4 gives the assumed 2-10 keV energy flux levels in  units of $\left[\rm 10^{-11}\rm \, erg \, cm^{-2}\, s^{-1} \right]$.
A flux of 1~mCrab corresponds to a 2-10 keV energy flux of 2.4$\times 10^{-11}$ ergs cm$^{-2}$ s$^{-1}$.
References: 
(1) Angel \& Stockman 1980, ARAA, 321,
(2) Tosti et al.\ 1998, A\&A, 339, 41,
(3) Jannuzi et al.\ 1993, ApJS, 85,265,
(4) Scarpa \& Falomo 1997, A\&A, 325, 109}
\end{table}
\subsection{Study of the Synchrotron X-ray Emission from AGN Jets}
The most likely emission mechanism responsible for X-ray signals from
HSP blazar jets is synchrotron radiation.  
The polarization of radio and optical synchrotron emission has been used to study the properties of AGNs for decades.  
Non-thermal emission tends to exhibit larger polarization degrees than thermal emission.
A power-law population of quasi-isotropic relativistic electrons
(with Lorentz factors $\gamma_e\gg 1$) with a differential distribution
$n_e(\gamma_e)\propto \gamma_e^{-p}$ in the {\it jet frame} generates a
degree of polarization $P_{\rm S}$ of
\begin{equation}
  P_{\rm S} \; =\; \frac{p+1}{p+7/3}\, \times \, 100\%
\label{eq:synch_pol_uniformB}
\end{equation}
in a uniform magnetic field \citep[e.g.][]{Ginz:65,Beke:66,Rybi:86}. 
This result is realized for a wide variety of pitch angle distributions of the electrons.
The basic exception is when the pitch angle lies 
within the small Lorentz cone of angle $1/\gamma_e$ with respect 
to the field direction. Then, the Doppler boosting of the dipolar 
radiation antenna pattern is distinctly different, and a separate 
regime of small-angle synchrotron emission \citep{Epst:73} is realized.  
This regime, perhaps less likely to be found in jets containing 
shocks or strong field turbulence, is addressed briefly below. 
We note that Eq.~(\ref{eq:synch_pol_uniformB}) generally applies 
also in the {\it observer's frame}, since the polarization degree
is Lorentz invariant. However, the polarization angle is affected 
by the Lorentz transformation and can result in its rotation 
between the two frames.

If the X-ray polarimetric observations localize regions of the jet with
quasi-uniform magnetic fields, this basic result yields a one-to-one
correspondence between the differential photon spectral index $\Gamma$
such that $n_{\gamma}(E) \propto E^{-\Gamma}$, and $P_{\rm S}$, since
$\Gamma = (p+1)/2$.   Hence, simultaneous measurements of
polarization and flux in distinct energy bins (at least 4) permit
spectropolarimetric probes of the field configuration. In the ideal case
described by Eq.~(\ref{eq:synch_pol_uniformB}), flat electron
distributions ($p=0$) yield $\Gamma = 1/2$ and $P_{\rm S} = 3/7$, already a strong
polarization signal.  At the other extreme, steep spectra (i.e.
turnovers with effective $p\gg 1$ and $\Gamma\gg 1$) yield polarization
degrees near 100\%.  The
principal hallmarks of this idealized, uniform (and laminar) 
field case include $>20\%$ polarization that is not strongly variable with flux levels over
time.  In contrast, the polarization angle $\phi$ can vary in an
extended time sampling of blazar X-ray emission (e.g. see Figure 1), in
either an organized fashion, such as a sinusoidal sweep expected for
helical jet structure, or in a more chaotic manner, corresponding to
more turbulent global field morphology within the jet.

A more probable scenario is that the observations only generate
good photon count statistics when integrating over blazar source volumes
that possess more complicated field morphologies.  Then field
tangling depolarizes the signal, perhaps down to 5-15\% levels, and
there is no coupling between spectral index $\Gamma$ and polarization
degree $P_{\rm s}$.  This reduces the diagnostic capability, but does not
eliminate it.  Since, within the narrow X-ray window, the effective
sampling of turbulent fields by radiating electrons is approximately
achromatic, one anticipates the same correlation as in uniform fields:
higher polarization degrees are associated with steeper spectra.  The
actual value of $P_{\rm s}$ will be a convolution over the integrated field
structure.  While this can be modeled with subjective choices, it will
be difficult to deconvolve observations to unambiguously constrain the
field components unless $P_{\rm s}$ exceeds 30\%.  However, the polarization
angle sweeps $\phi (t)$ with time will probably disentangle static,
uniform field structure environments from those where the large scale
average magnetic field vector displays an organized, but non-uniform
(e.g. helical) geometry.  We note also that the signature of viewing
perspectives down the mean magnetic field line is low linear
polarization coupled with rapidly variable $\phi$, i.e. approximating
circular polarization.  This maps over to large/rapid polarization angle
swings in time traces, such as those detected in radio pulsars.

The source candidates of this science investigation are identical to those
of the previous science investigation. The candidate sources are listed in
Table \ref{tflux}.
\subsection{Identify the Dominant Target Photon Population 
for the Inverse Compton Emission from LSPs, ISPs, and FSRQs}
For LSPs, some ISPs, and FSRQs the soft X-ray observations sample the inverse Compton component.
The GEMS observations should be able to decide between a Synchrotron Self Compton (SSC) and an 
External Compton (EC) origin of the inverse Compton emission \citep{Kraw:11a,Kraw:11b}. 

In the case that the X-ray emission is of SSC origin, the X-ray inverse Compton emission 
should exhibit similar polarization properties (polarization degree and polarization 
direction) as the optical synchrotron emission \citep{Pout:94,Celo:94,Kraw:11b}. 
If the X-ray emission is of EC origin, the X-ray polarization degree is 
expected to be lower ($\ll$10\%) than the optical polarization degree \citep{Pout:94,Kraw:11b}. 

\begin{figure}[tb]
\begin{center}
\includegraphics[width=3.5in]{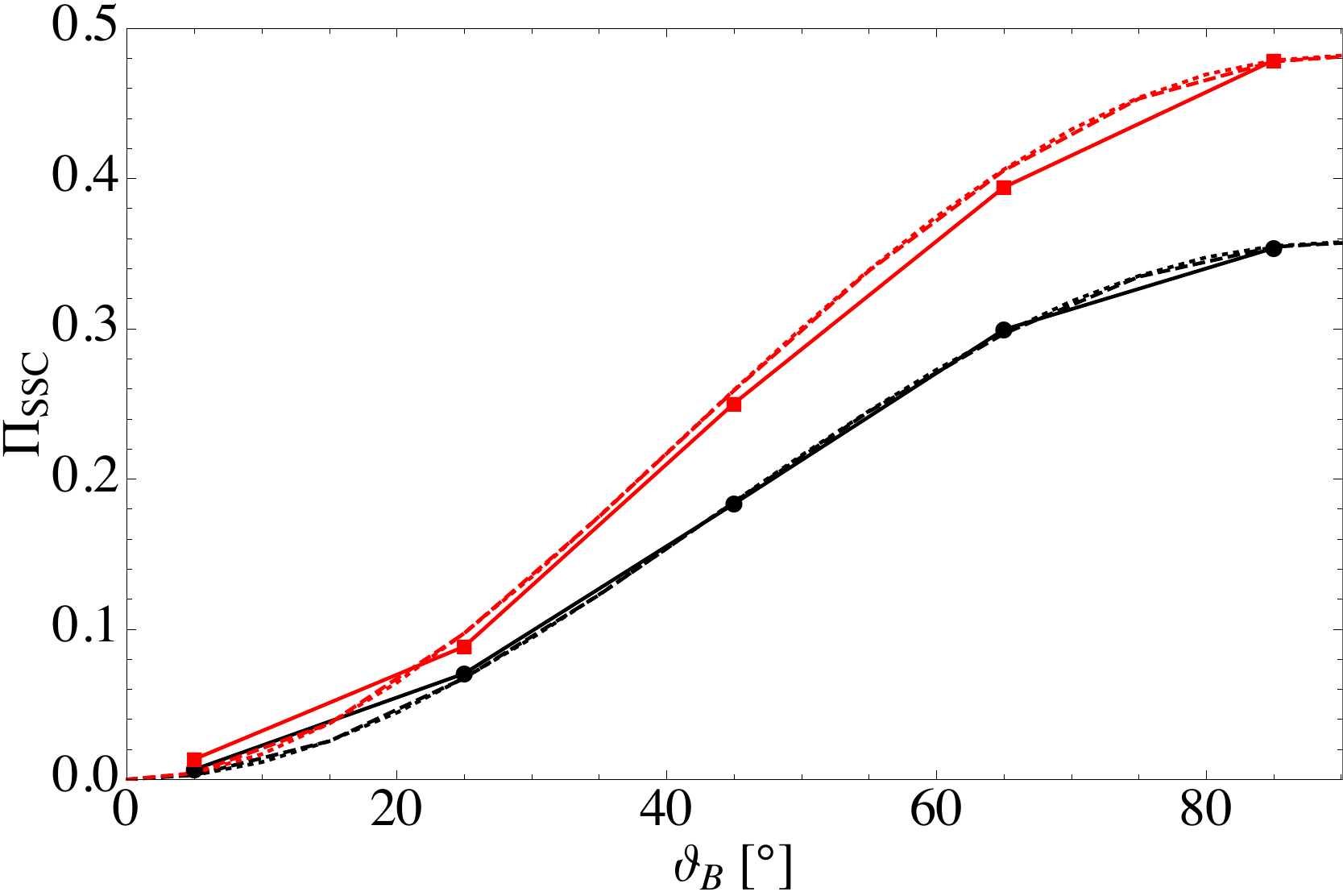}
\caption{\label{yp} Polarization degree of the SSC emission as function of the 
angle $\vartheta_{\rm B}$ between the magnetic field and the line of sight for 
a spectral index $\alpha=1$ of the target photons, and a differential electron number index 
$p=3$ (squares), as well as for $\alpha=0.5$ and $p=2$ (circles). The lines show various
semi-analytic approximation to the polarization degree \citep[see][]{Kraw:11b}.
Results are shown in the ``central power law region of the SSC emission'', 
at energies well above and well below the low-energy and high-energy 
cutoffs of the emission, respectively.}
\end{center}
\end{figure}
The polarization degree of the SSC emission $P_{\rm SSC}$ can be a sizable fraction 
of that of the synchrotron emission $P_{\rm S}$ (see Fig.\ \ref{yp}) 
depending on the  angle $\vartheta_{\rm B}$ between the magnetic field direction 
and the line of sight.
Given that the synchrotron polarization degrees are usually rather high 
(close to the most optimistic theoretical predictions), the SSC model definitively 
predicts a high degree of polarization in the GEMS energy band.

The observations should target a few ($\sim$ 5) sources of the three different 
source classes LSPs, ISPs, and FSRQs to identify for each of these classes the dominant
seed photon population.

Table \ref{t2} lists seven source candidates. The sources were the brightest 
BeppoSAX LSPs and FSRQs in the MECS (2-10 keV) energy band \citep{Dona:05}.
All except 3C 273 show high degrees of polarization. For the source S52116+81
we did not find published optical polarization information.
\begin{table}[tb]
\begin{center}
\begin{tabular}{|p{2.4cm}|p{1.2cm}|p{1.2cm}|p{2.4cm}|p{2.4cm}|}
\hline
Object			& Class & $z$   & $F(2-10\,\rm keV)$ & opt.\ pol $\left[ \% \right]$ \\ \hline \hline
BL Lac      & LSP   & 0.069 & 2         & 2-23 (1)		\\
ON 231		& LSP   & 0.102 & 0.4       & 2-10 (1)		\\
OJ 287		& LSP   & 0.306 & 0.13-0.24 & 1-40 (2)		\\
3C 273      & FSRQ  & 0.158 & 9-11		& 1 (3)			\\
3C 279      & FSRQ  & 0.586 & 1         & 2-43 (4,5)	\\
S52116+81   & FSRQ  & 0.084 & 1.6       & nm			\\
1ES0836+710 & FSRQ  & 2.172 & 3         & 15 (5)		\\
\hline
\end{tabular}
\end{center}
\caption{\label{t2}  \small Target list for the science investigation  
``Identify the Dominant Target Photon Population 
for the Inverse Compton Emission from LSPs, ISPs, and FSRQs''.
Column 4 gives the assumed 2-10 keV energy flux levels in  units of $\left[\rm 10^{-11}\rm \, erg \, cm^{-2}\, s^{-1} \right]$.
A flux of 1~mCrab corresponds to a 2-10 keV energy flux of 2.4$\times 10^{-11}$ ergs cm$^{-2}$ s$^{-1}$.
References:
(1) Angel \& Stockman 1980, ARAA, 8, 321,
(2) Efimov et al.\ 2002, A\&A, 381, 408,
(3) Impey et al. 1989, Valtaoja et al. 1990, Valtaoja et al. 1991,
(4) Marscher et al.\ 2008, Nature, 452, 966,
(5) Scarpa \& Falomo 1997, A\&A, 325, 109}
\end{table}
\subsection{Closeup view of a misaligned blazar - the case for GEMS observations of Cen-A}
At a distance of 3.4 $\pm$0.3 Mpc, Centaurus A is the nearest and best studied radio galaxy. 
As such, it provides us with an ideal laboratory to examine the physics of relativistic jets, 
their connection with the accretion flow in the center of AGN, and unification models for 
radio-loud sources. In the radio it shows structures on all resolved scales, 
starting with the giant radio lobes and two-sided jets that subtend $\sim$10$^{\circ}$ on 
the sky, oriented primarily Northeast-Southwest. Based on radio morphology and power it is 
classified as an FRI.

The host galaxy, NGC5128, is a giant elliptical containing a kpc-scale dust lane 
which obscures the center. Optical spectra of the nucleus show no evidence for 
broad emission lines, and Cen A is optically classified as a Narrow Line Radio Galaxy. 
As the brightest X-ray emitting active galaxy in the sky, \citet{Chia:01} used RXTE
data to construct the spectral energy distribution from the radio band through
about 100 MeV.  The peak of $\nu F_{\nu}$ is $\sim 10^{42}$ erg s$^{-1}$ at about 100
keV and the luminosity in the 2-6 keV band is about $3 \times 10^{41}$ erg~s$^{-1}$.

There are conflicting reports of X-ray emission from the giant lobes but the most
reliable of these \citep{1981ApJ...245..840M} indicate that there is less than
$3 \times 10^{-11}$ erg~s$^{-1}$~cm$^{-2}$ from the giant lobes, or $< 4 \times
10^{40}$ erg s$^{-1}$.   However, these lobes are very diffuse and are well
outside the GEMS field of view when the core is targeted.
% add ref. to figure here: CenA.pdf
The ``inner'' jet and the counterjet, extending several
arcmin from the core, are detected in ultra-deep Chandra 
images (Hardcastle et al. 2009), allowing detailed studies of particle 
acceleration and shocks. The total power in the resolved emission is
of order $3 \times 10^{40}$ erg~s$^{-1}$, or about 10\% of the power from the
core.
Chandra also shows the presence of several hundreds 
X-ray binaries scattered in the halo of the galaxy \cite{2009ApJ...698.2036K}. 
Integrating the luminosity function of the binaries gives an integrated luminosity
of $\sim 10^{39}$ erg~s$^{-1}$, so the binaries contribute less than 1\% of the
total flux.

The nuclear emission dominates at CCD resolution. A recent Suzaku 
observation (Markowitz et al. 2007) detects the source up to 250 keV
and obtain a 2-10 keV luminosity of $6.7 \times 10^{41}$ erg~s$^{-1}$.
The hard X-rays are fit by two power laws of the same slope, 
absorbed by columns of 1.5 and 7$\times$10$^{23}$ cm$^{-2}$, respectively. 
The spectrum is consistent with previous suggestions that the 
power-law components are X-ray emission from the subparsec VLBI jet 
and from Bondi accretion at the core, but it is also consistent with 
a partial-covering interpretation. Below 2 keV, the spectrum is dominated by 
thermal emission from the diffuse plasma and is fit well by a 
two-temperature VAPEC model, plus a third power-law component 
to account for scattered nuclear emission, jet emission, and 
emission from X-ray binaries and other point sources. 
Narrow fluorescent emission lines from Fe, Si, S, Ar, Ca, and Ni are detected. 
The Fe K$\alpha$ line width yields a 200 light day lower limit on the distance 
from the black hole to the line-emitting gas. 
The thermal components are very soft

Fermi recently detected Centaurus A at GeV energies
\citep{2010ApJ...719.1433A}, confirming the previous 
EGRET detection, and by HESS at TeV. The Spectral Energy Distribution from 
radio to $\gamma$-rays can be reasonably well fitted by a Òmisoriented blazarÓ 
model Ð synchrotron (radio to optical) + SSC (X-rays to GeV) 
from a one-zone model \citep{Chia:01}; however, the non-simultaneous TeV 
data are not consistent with the extrapolation from the GeV, lying above it 
by almost one order of magnitude. This has been interpreted in terms of a spine-sheath jet 
model with a bulk Lorentz factor of $\Gamma = 2$ in the sheath.
As this component will dominate in the soft X-ray band and SSC models generally
predict strong polarization (unlike external Comptonization of uniform emission)
a detection is deemed likely to provide a very good datum as input for
SSC models.  The spatial scale will likely be much smaller than even probed
on VLBI scales, so the polarization direction should give the orientation of the jet at
sub-mas angular scales.  Due to variability of a factor of a few, it would be very
useful to constrain the high energy portion of the SSC component with NuSTAR
or comparable data.
\subsection{Search for Lorentz Invariance Violations}
For the last two decades theoretical studies and experimental searches of Lorentz 
Invariance Violation (LIV) have received a lot of attention \citep[see the reviews by][]{Matt:05,Jaco:06,Will:06}.
On general grounds one expects that the two fundamental theories of our time
the General Theory of Relativity and the Standard Model of Particle Physics
can be unified at the Planck energy scale. Under certain conditions deviations 
from the two theories may be observable even at much lower energies, 
e.g.\ if effects such as a tiny difference between the propagation 
speed of orthogonally polarized photons accumulate over cosmological 
distances to become measurable \citep{Coll:97,Coll:98}.

Possible consequences of LIV are energy and helicity dependent photon 
propagation velocities. The energy dependence can be constrained by 
recording the arrival times of photons of different energies emitted 
by distant objects at approximately the same time \citep{Amel:98}, 
e.g.\ during a Gamma-Ray Burst \citep{Abdo:09}
or a flare of an Active Galactic Nucleus \citep{Ahar:08}. 
The energy and helicity dependence can be constrained by measuring 
how the polarization direction of an X-ray beam of cosmological 
origin changes as function of energy \citep{Gamb:99}.

The detection of a non-zero polarization from a cosmological source
at 2 keV and 10 keV constrains the phase difference with an accuracy of $\Delta\phi\,<$ 
$\sim\pi$ and would allow us to access group velocity differences down to
\begin{equation}
\frac{\Delta v_{\rm g}}{c} \,\approx\,2\,
\Delta\phi (L/\lambda)^{-1} \,\approx\, 10^{-35}\,
\frac{\Delta \phi}{\pi}
\left(\frac{E_{\gamma}}{\rm 5\,keV}\right)^{-1} 
\left(\frac{L}{\rm 1\,Gpc}\right)^{-1} 
\label{LIVeq}
\end{equation}
where $E_{\gamma}$ and $\lambda$ are the mean energy and wavelength of the observed 
photons. 

Equation (\ref{LIVeq}) shows that deviations of the speed of light
can be measured with an accuracy proportional to $(L/\lambda)^{-1}$ 
$\propto\,E_{\gamma}^{-1}$. Measurements at higher energies and shorter 
wavelengths thus lead to better constraints.
LIV is thought to be a high-energy phenomenon; the deviation of the
speed of light is expected to increase with energy.
Using an expansion in powers of $E_{\rm \gamma}$
\begin{equation}
\frac{\Delta v_{\rm g}}{c}\,=\,\sum_{n=1}^{\infty} \eta^{(n)} E_{\gamma}^{\,\,n}
\end{equation}
we infer that the accuracies of the birefringence constraints 
on the coefficients $\eta^{(n)}$ scale as $E_{\gamma}^{-(n+1)}$.
Polarization measurements at X-ray and $\gamma$-ray energies enable extremely sensitive tests of Lorentz 
invariance violating models that predict a helicity dependence of the speed of light \cite[e.g.][]{Gamb:99,Kaar:04,Fan:07,Kost:13}. 
General constraints on the Lorentz violating terms of effective field theories beyond the standard model
require constraints form a sample of sources \citep{Kost:09,Kost:13}. A GEMS-like mission could deliver 
such constraints for a sample of AGNs.
\section{Proposed Observation Program} 
\label{strategy}
% 1000 ksec, 1 mCrab, 2%
\begin{table}[t]
\begin{center}
\begin{tabular}{|p{2.4cm}|p{1.2cm}|p{1.2cm}|p{2.4cm}|p{2.9cm}|p{2.4cm}|}
\hline
Object			& Class & $z$   & $F(2-10\,\rm keV)$ & Obs. Time $\left[\rm ksec \right]$ & MDP $\left[\rm \% \right]$ \\ \hline
\hline \multicolumn{6}{| c |}{\textbf{Snapshot Survey}} \\ \hline
Cen A       & RG  & 0.0018  & 13        & 20 & 6.7 \\
Mrk 421			& HSP   & 0.031 & 48    & 20 & 5.4 \\ 
Mrk 501			& HSP   & 0.034 & 24    & 20 & 7.7 \\
1ES 1959+650	& HSP	& 0.047 & 4.8	& 20 & 17.3 \\
PKS 2155-314	& HSP   & 0.116 & 12	& 20 & 10.9 \\
1ES 1218+304    & HSP	& 0.184 & 24   & 20 & 7.7  \\ 
BL Lac      & LSP   & 0.069 & 2         & 20 & 26.8 \\
3C 273      & FSRQ  & 0.158 & 9-11		& 20 & 12.0 \\
S52116+81   & FSRQ  & 0.084 & 1.6       & 20 & 30.1 \\
1ES0836+710 & FSRQ  & 2.172 & 3         & 20 & 21.9 \\
\hline \multicolumn{6}{| c |}{\textbf{Target of Opportunity Observations}} \\ \hline
tbd			& -   & - & -    & 200 & - \\ 
tbd			& -   & - & -    & 200 & - \\ 
\hline \multicolumn{6}{| c |}{\textbf{Deep Observations of Key Sources}} \\ \hline
Mrk 421			& HSP   & 0.031 & 48    & 600 & 0.8 \\ 
Cen A			& HSP   & 0.031 & 13    & 200 & 1.8 \\ 
1ES0836+710     & FSRQ  & 2.172 & 3     & 600 & 3.3 \\ \hline
\end{tabular}
\end{center}
\caption{\label{program}  \small Proposed observation program. The total observation time is 2,000 ksec.
Column 4 gives the assumed 2-10 keV energy flux levels in  units of $\left[\rm 10^{-11}\rm \, erg \, cm^{-2}\, s^{-1} \right]$.
A flux of 1~mCrab corresponds to a 2-10 keV energy flux of 2.4$\times 10^{-11}$ ergs cm$^{-2}$ s$^{-1}$.}
\end{table}
A tentative observation program for a 9-month GEMS-like mission could look as follows:
\begin{enumerate}
\item We envision a snapshot survey with relatively short ($\sim$20 ksec) exposures at the 
beginning of the 9-month observation window. For each of the six science investigations, 
we envision to observe all the candidate sources listed in Table \ref{tflux} and \ref{t2} plus Cen A.
For 10 mCrab sources, the observations would allow us to detect polarization degrees down to 2\%.
The observations could establish a baseline for later observations.
For science investigations \#1 and \#2 the observations would show which sources exhibit
high degrees of polarization.
\item The initial snapshot survey should be combined with a alert-driven target of opportunity (ToO) program.
If a sources goes into a bright flaring phase it could be observed for 5 days with 40 ksec 
observation time each day.
\item In addition to the snapshot survey and the alert-driven  ToO program, we recommend deep observations
of three key sources, i.e.\ Mrk 421, Cen-A, and 1ES0836+710. The Mrk 421 observations will be used
to carry through the science investigation \#1 and \#2. Cen-A is a unique target owing to its proximity 
and its X-ray brightness; it would allow us to address science investigation \#4. 
1ES 0836+710 is an attractive target for science 
investigation \#3 and \#5.
\end{enumerate}

\section{Epilogue} 
\label{gems}
GEMS was proposed in response to the NASA SMEX announcement of opportunity  in December 2008, was selected for phase A development in 2009 and down selected for phase B in 2010. A technically successful Preliminary Design Review  was held in Feb 2012. NASA Science Mission Directorate (SMD) indicated their intention to non-confirm (or cancel) in May 2012; the SMD decision was based on concerns that the eventual cost would be too high.\\[2ex]

{\it Acknowledgments:} HK acknowledges support from NASA (grant NNX10AJ56G), and the Office of High Energy Physics of the US Department of Energy. 

\end{document}